\documentclass{vldb}

\usepackage{cite}
\usepackage{booktabs} 
\usepackage{graphicx}
\usepackage{balance}
\usepackage{pifont} 
\usepackage{longtable}
\usepackage{enumitem} 
\usepackage[algoruled, linesnumbered, vlined]{algorithm2e}
\usepackage{algorithmic}
\usepackage{setspace}
\usepackage{multirow}
\usepackage{url}
\usepackage[T1]{fontenc}

\usepackage{sparklines}
\usepackage{xspace}
\usepackage{xcolor}
\usepackage{soul}
\usepackage{marginnote}
\newcommand{\ie}{i.e.,\ }
\newcommand{\eg}{e.g.,\ }

\newcommand{\tinyskip}{\vspace{3pt}}
\newcommand{\mypar}[1]{\tinyskip\noindent\textbf{#1.}\xspace}

\newcommand{\F}{\mbox{Fig.\hspace{0.25em}}}

\newenvironment{myitemize}{%
\begin{itemize}[leftmargin=1em, itemsep=.1em, parsep=.1em, topsep=.1em,
    partopsep=.1em]}
{\end{itemize}}

\newenvironment{myenumerate}{%
\begin{enumerate}[leftmargin=1em, itemsep=.1em, parsep=.1em, topsep=.1em,
    partopsep=.1em]}
{\end{enumerate}}

\newenvironment{structure*}{\color{blue}\begin{myenumerate}}{\end{myenumerate}}

\sethlcolor{yellow}
\newcommand{\update}[3][0em]{#3}

\vldbTitle{Data Market Platforms: Trading Data Assets to Solve Data Problems}
\vldbAuthors{Raul Castro Fernandez, Pranav Subramaniam, Michael J. Franklin}
\vldbDOI{https://doi.org/10.14778/3407790.3407800}
\vldbVolume{13}
\vldbNumber{11}
\vldbYear{2020}

\begin{document}

\title{Data Market Platforms: Trading Data Assets \\to Solve Data Problems}

\numberofauthors{1}

\author{
\alignauthor 
Raul Castro Fernandez, Pranav Subramaniam, Michael J. Franklin\\
\affaddr{The University of Chicago}\\
\email{[raulcf,psubramaniam,mjfranklin]@uchicago.edu}
\alignauthor 
%
%
}

\maketitle

\begin{abstract}

Data only generates value for a few organizations with expertise and resources
to make data shareable, discoverable, and easy to integrate. Sharing data that
is easy to discover and integrate is hard because data owners lack
\emph{information} (who needs what data) and they do not have \emph{incentives}
to prepare the data in a way that is easy to consume by others. 

In this paper, we propose data market platforms to address the lack of
information and incentives and tackle the problems of data sharing, discovery,
and integration. In a data market platform, data owners want to \emph{share}
data because they will be rewarded if they do so. Consumers are encouraged to
share their data needs because the market will solve the \emph{discovery} and
\emph{integration} problem for them in exchange for some form of currency.

We consider internal markets that operate within organizations to bring down
data silos, as well as external markets that operate across organizations to
increase the value of data for everybody. We outline a research agenda that
revolves around two problems. The problem of market design, or how to design
rules that lead to desired outcomes, and the systems problem, how to
implement the market and enforce the rules. Treating data as a first-class asset
is sorely needed to extend the value of data to more organizations, and we
propose data market platforms as one mechanism to achieve this goal.

\end{abstract}

\section{Introduction}
\label{sec:introduction}

Data is the new oil~\cite{dataoil-economist}. Like oil, it generates enormous
value for the individuals and organizations that know how to tap into and refine
it. Like oil, there are only a select few who know how to exploit it. In this
paper, we present our vision for data market platforms: collections of
\emph{protocols} and \emph{systems} that together enable participants to exploit
the value of data.

Data only generates value for a few organizations with expertise and resources
to solve the problems of data \emph{sharing}, \emph{discovery}, and
\emph{integration}. These problems remain difficult despite the many
contributions of the database community (and others) to theory, algorithms, and
systems. Solving these problems precedes advanced analytics and machine
learning, explaining why the large majority of organizations only
partially benefit from the data they own.


The central argument of this paper is that sharing, discovering, and integrating
data is hard because data owners lack \emph{information} and \emph{incentives}
to make their data available in a way that increases consumers' utility. Data
owners do not know what data consumers want and in what format, and they are
disincentivized to share data that may leak confidential information. Even within
organizations, sharing data is time-consuming and it is not clear what is the
return on investment. \update{R4D6a}{When data is finally shared via open data
portals}~\cite{datagov, datagov2}, \update{}{or via data lakes}~\cite{datalake1, datalake2},
\update{}{consumers still need to discover data that satisfies their needs and
integrate it into the format they want, which is typically different than the
format in which they found the data.}


\emph{Data market platforms} establish rules to share data in a way that is easy
to discover and integrate into the format consumers need because markets
understand the data consumers need and communicate this need to the owners.  In
a data market platform, data owners are encouraged to share their data because
they may receive profit if a consumer is willing to pay for it.  Consumers are
encouraged to share their data needs because the market will solve the discovery
and integration problems for them in exchange for some reward \eg money. By spreading
information among interested parties and incentivizing them, data market
platforms bring data value to all participants.

In recent years we have seen the appearance of many data markets such as
Dawex~\cite{dawex}, Xignite~\cite{mkt2}, WorldQuant~\cite{mkt2} and others to
directly buy and sell data. Data brokers~\cite{databroker1, databroker2} are
active participants in the Internet economy by trading user data for ads. Many
of us participate in the data economy when we exchange our personal data for
Internet services~\cite{inetservice1, inetservice2}. And hospitals and other
health institutions have started exchanging data to improve patient care and
treatments. The interest in trading data is not new. Economists have been
considering these problems for decades~\cite{varian1, valuedata1} and the
database community has made progress in issues such as pricing queries under
different scenarios~\cite{qbdp15, chenMLQ19, revMax19}. We believe the time is
ripe to \textbf{design and implement data market platforms that tackle the
sharing, discovery, and integration problems}, and we think the database
community is in an advantageous position to apply decades of data management
knowledge to the challenges these new data platforms introduce. In this paper,
we outline challenges and a research agenda around the construction of data
market platforms. 

\subsection*{Using Data Markets to Solve Data Problems}

Consider a market with \emph{sellers}, \emph{buyers}, and an \emph{arbiter} that
facilitates transactions between buyers and sellers. 
Sellers and buyers can be individuals, teams, divisions, or whole
organizations. Consider the following example: 

\begin{myitemize}
\item Buyer $b_1$ wants to build a machine learning classifier and needs
features $\langle a, b, d,
e\rangle$,
and at least an accuracy of 80\% for the responsible engineer to trust the
classifier.
\item Seller 1 owns a dataset $s_1 = \langle a, b, c\rangle$ that they want to share with the
arbiter.
\item Seller 2 owns a dataset $s_2 = \langle a, b', f(d)\rangle$ that they will not share with the
arbiter unless the dataset is guaranteed to not leak any business information.
\end{myitemize}

\mypar{Details of the example} In the example $f(d)$ is a function of $d$, such
as a transformation from Celsius to Fahrenheit. The function can also be
non-invertible, such as a mapping of employees to IDs. Note that neither $s_1$
nor $s_2$ owns attribute $e$, which $b_1$ wants: we discuss this attribute in
Section~\ref{subsec:opportunities}. Last, $b'$ is an attribute similar to $b$.

\textbf{Challenge-1.} Using $s_1$ \update{R3}{alone is not enough to satisfy}
$b_1$'s \update{}{needs.
The arbiter must incentivize Seller 2 to share} $s_2$. \update{}{The first challenge is to
compensate sellers so they are incentivized to share their data. This requires}
\emph{pricing} and \emph{revenue allocation} mechanisms.

\textbf{Challenge-2.} \update{R3}{Without knowing how useful} $s_1$ or $s_2$
\update{}{will be to train
the ML model,} $b_1$ \update{}{risks overpaying for the datasets. The second challenge is
to guarantee} $b_1$ \update{}{a certain quality before paying for the datasets.}

\textbf{Challenge-3.} Neither $s_1$ nor $s_2$ \emph{alone} fulfill $b_1$'s need.
$b_1$ \update{R3}{may not want to pay for two incomplete datasets that still need to go
through a slow and expensive integration process. 
The third
challenge requires combining the datasets supplied by sellers to satisfy buyers'
needs. This combination can be arbitrarily complex, such as determining how to
go from} $f(d)$ to $d$, which is the attribute the buyer wants.


\textbf{Challenge-4.} \update{R3}{The degree of trust between sellers, buyers,
and arbiter may vary. While it is conceivable that participants within an
organization trust each other, this may not be the case in external markets. The
fourth challenge is to help all participants trust each other}.

\begin{figure}[t]
  \centering
  \includegraphics[width=0.85\columnwidth]{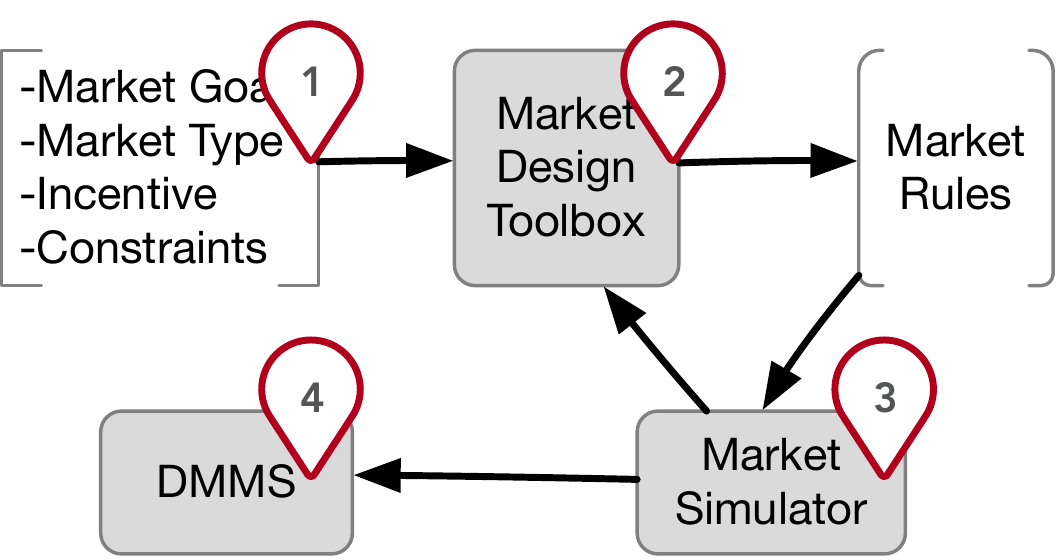}
\caption{
Given a market definition (1), a market design toolbox (2) generates the market
rules, which are simulated in (3), possibly refined, and finally deployed on a
DMMS (4).
}
\label{fig:marketdesigntimeline}
\end{figure}

\subsection*{Requirements of Data Market Platforms}

The previous challenges motivate a list of requirements for data market
platforms:


\mypar{Value of Data} The market must price datasets so it creates a demand of
buyers requesting data and a supply of datasets shared by sellers.


\mypar{Market Design} Without the right rules to govern the participation of
sellers, buyers, and arbiter, the market will be gamed and collapse. The key
intellectual questions in this area are on how to design the market rules 
when the asset is \emph{data}, which is freely replicable and can
be combined in many different ways~\cite{imf}. A requirement of a
data market platform is to be resilient to strategic participants. 



\mypar{Plug n' Play Market Mechanisms} Markets can be of many \emph{types}: i)
internal to an organization to bring down silos of data, in which case employee
compensation may be bonus points; ii) external across companies where money is
an appropriate incentive; iii) across organizations but using the shared data as
the incentive, such as hospitals exchanging medical data~\cite{hospital1}. 
The \emph{goals} of the market may also be
varied, from optimizing the number of transactions to social welfare, data utility,
and others. A requirement of data market platforms is to flexibly support
markets of different types and with different goals. 


\mypar{Arbiter Platform} Because the supplied data will have a
different format than the demanded data, a key requirement to enable
transactions between sellers and buyers is an arbiter platform that can combine
datasets into what we call \emph{data mashups} to match supply and demand. 

In the example above the arbiter can combine $s_1$ and $s_2$'s
datasets, obtaining a dataset that is much closer to $b_1$'s need. For that, it
needs to understand how to join both datasets and needs to find an inverse
mapping function $f'$ that would transform $f(d)$ into $d$ if such a function
exists, or otherwise find a mapping table that links values of $f(d)$ to values
of $d$. In addition to these relational and non-relational operations the
arbiter must also support \emph{data fusion} operators to contrast different
sources of data of the same topic. Briefly, the data fusion operators we
envision produce relations that break the first normal form, that is, each cell
value may be multi-valued, with each value coming from a differing source. Data
fusion operators are appropriate when buyers want to contrast different sources
of information that contribute the same data, \ie weather forecast signals coming from a
city dataset, a sensor, and a phone. As an illustration, note Seller 2 owns
attribute $b'$ which is almost identical to $b$, but has some non-overlapping,
conflicting information. A buyer may be interested in looking at both signals,
or at their difference, or at their similarities, etc. A data mashup is a
combination of datasets using relational, non-relational, and fusion operations.



\mypar{Data Market Management System (DMMS)} In addition to the arbiter platform, a
data market platform calls for platforms to support sellers and buyers.
\update{R2W1}{Sellers need access to statistical privacy techniques so they feel confident
when sharing data. For example, without the capability of dealing with PII
information, Seller 2 will not share data despite the potential monetary benefit of
doing so because leaking such information may be ilegal}. Buyers need to have
the ability to describe with fine granularity their data needs and the money
they are willing to pay for a certain \emph{degree of satisfaction} achieved on
a given task. In the example above, the buyer should have the ability to define
that they are only willing to pay money for a classifier that achieves at least
80\% accuracy. A requirement of a data market management system is to offer
support for sellers, buyers, and the arbiter.


\mypar{Building Trust} The degree of trust among sellers, buyers, and arbiter
will differ depending on the scenario, \ie whether in an internal market or
across the economy. A requirement of a DMMS is to implement mechanisms to help
participants trust each other, such as using decentralized
architectures~\cite{blockchain2}, implementing computation over encrypted
data~\cite{processingencrypteddata}, and supporting contextual
integrity~\cite{contextualintegrity}.


\mypar{Market Simulator} The mathematics used to make sound market designs do
not account for evil, ignorant, and adversarial behavior, which exists in
practice. For that reason, it is necessary to simulate market designs under
adversarial scenarios before their deployment. Hence, a data market platform
calls for a market simulator.

\subsection*{Our Vision}

In the remainder of this paper, we delve into the details of our vision for data
markets. \update{R4D7}{The challenges are wide ranging and many fall directly within the
territory of the database community. We outline our own strategy to
tackle them in Section}~\ref{subsec:builddmms}.

Our vision is to produce market designs for different scenarios (points (1) and
(2) in \F\ref{fig:marketdesigntimeline}) using a market design toolbox
(Section~\ref{sec:design}). The market design toolbox uses techniques from game
theory and mechanism design~\cite{mechanismdesign} to deal with the modeling and
engineering of rules in strategic settings, such as data markets. Every market
design is tested using a data market simulator (point (3))
(Section~\ref{sec:evaluation}), before being finally deployed in a DMMS (point
(4)), presented in Sections~\ref{sec:platform} and \ref{sec:mashup}. While the
output of the market design toolbox is a collection of equations, the output of
the DMMS is software. There is an explicit interplay between market design and
DMMS that constrains and informs the capabilities of the other. Exploring such
an interplay is a critical aspect of our proposal. Before delving into the
details of engineering market designs we discuss briefly the value of data
(Section~\ref{sec:datavalue}). The paper concludes with a discussion of the
impact of data markets~\ref{sec:impact}, related work~\ref{sec:relatedwork}, and
conclusions~\ref{sec:discussion}.

%

\section{The Value of Data}
\label{sec:datavalue}


What's the value of data? This question has kept academics and practitioners in
economics, law, business, computer science and other disciplines
busy~\cite{varian1, valuedata1, imf}. \update{R3, R4D9}{This question has been explored from a
macroeconomic angle to study the impact of data in an econom}y~\cite{macro1,
macro2}. \update{}{It has been studied from a microeconomic perspective to study the
impact on individual companies and firms}~\cite{bigdatafirm18, predwbig13}. \update{}{Answering the
question in its full generality is outside the scope of this paper. Instead, we
focus on the narrower yet challenging goal of choosing a price that satisfies
sellers and buyers.}

The crux of the problem is that the value of a dataset may be different for a
seller and a prospective buyer. Sellers may choose to price their datasets based
on the effort they spent in acquiring and preparing the data, for example.
Buyers may be willing to pay for a dataset based on the expectation of profit
the dataset may bring them: \eg how much they will improve a process and how
valuable that is. None of these strategies is guaranteed to converge to a
price---and hence a transaction agreement---between participants. 

And yet, this is how prices are set in current markets of datasets such as
Dawex~\cite{dawex}, Snowflake's Data Exchange~\cite{snowflake-exchange}, and
many others. Sellers choose a price for datasets without knowing buyers'
valuations and buyers who are willing to pay that price obtain the datasets,
without knowing how useful the dataset is to solve their problem. This leaves
both sellers and buyers unsatisfied. Buyers may pay a high price for datasets
that do not yield the expected results. Similarly, sellers may undervalue
datasets that could yield more profits because they lack information about what
buyers want.

\update{R3, R4D9}{Finally, beyond buyers' and sellers' opinions on the dataset
price, we must also take into consideration the
externalities that datasets create. First, if a dataset is exchanged with exclusive
rights (assuming that this is possible), the transaction creates an externality on
those entities that are denied access to the dataset. Second, datasets that
contain information about people generate an externality: the potential harm and
privacy loss. Last, a transaction involving a
dataset and a price communicates to others the value those firms put on that
data. According to economic theory, all these transaction costs have an impact
on real-world markets}~\cite{transactioncoase}, \update{}{and consequently they must be
taken into consideration.}

\update{}{In this vision paper, we constrain our agenda to finding a price
that permits establishing a transaction between seller and buyer, without
identifying the specific components of that price, i.e., externality. This is a
first step towards studying the} \emph{value of data}. Having set this goal, we
move on to discuss how to price a dataset.

\subsubsection*{Intrinsic vs Extrinsic Properties of Data}

A tempting option to the database community is to price datasets based on 
intrinsic properties, such as quality, freshness, whether the data include
provenance information or not, etc. Unfortunately, pricing datasets based on the
intrinsic properties of data \emph{alone} does not work. Consider the following
scenarios.

\mypar{Intrinsic properties alone are insufficient} \update{R3}{For example,
given datasets with different number of missing values that are otherwise
equivalent, we tend to think the one with fewer missing values is more
}\emph{valuable}. \update{}{However, the value of this dataset ultimately
depends on the specific task for which it is used: if both datasets solve the
task equally, the different ratio of missing values does not matter. Similarly,
fresher datasets are intuitively considered more valuable, but this is not true
if our analysis is concerned with some past date, in which case an older dataset
that corresponds to that date is more valuable.  Finally, the lessons from
feature engineering to train statistical and machine learning models tells us
that more data is not always better; diversity with respect to the task matters
as well.} Having established that intrinsic properties alone cannot be used to
set a dataset's price we discuss how to find such price.

In the markets we envision, the price of a dataset is set by the arbiter
based on the economic principles of supply and demand~\cite{supplyanddemand}. A
dataset that lots of buyers want will be priced higher than a 
dataset that is hardly ever requested, regardless of the intrinsic properties of
such datasets. In other words, the value of a dataset is primarily extrinsic. 

\mypar{The role of intrinsic properties} Intrinsic properties are important
insofar the buyers indicate a preference as part of their data demands. For
example, as a consequence of a buyer requesting a dataset with few missing
values, sellers who provide those datasets will profit more. However, intrinsic
properties do not have an associated value without explicit demand for them.


\section{Designing the Market Rules}
\label{sec:design}

\update{}{In this section we give a definition for} \emph{market design} in
Section~\ref{subsec:marketdesigndef}, \update{}{we explain the challenges of
engineering market designs for data (Section} \ref{subsec:mdchallenges}),
\update{}{and then discuss how different requirements call for different market
designs in Sectio}n~\ref{subsec:mdtypes}.  \update{}{Finally, we close the
section with a FAQ that briefly answers some questions we cannot cover here.}

\update{}{We start by giving a simple market model that we use throughout this section to
define and illustrate the ideas:}

\mypar{Market Model} \update{}{The market consists of buyers} ($b_i \in B$), sellers ($s_i
\in S$), and an arbiter, $a$. Sellers own datasets ($d_i$) \update{}{that they are willing
to share with the arbiter in exchange for money. Buyers want to obtain datasets
that solve their problems. In general, the datasets buyers need do not directly
match the datasets sellers offer. The arbiter's goal is to combine datasets
offered by sellers into a} \emph{mashup} $m$ \update{}{that fulfills buyers' needs. In exchange for
finding the mashup, buyers transfer an incentive (we use money as an incentive
throughout this section) to the arbiter. The arbiter uses the money from the
buyers to compensate the sellers who contributed datasets that were part of the
sold mashup.} 

\mypar{A brief discussion of buyers' utility in an external market}
\update{R4D4}{In its
simplest form, we assume a buyer's utility,} $u_i$, \update{}{is determined by a
function:} $u_i = v_i - p_i$, where $v_i$ \update{}{is the
private valuation the buyer assigns to the data and} $p_i$ \update{}{is the price they pay
for it. For the buyers to participate in the market in a way that does not
decrease their utility, they must know} $v_i$. \update{}{In the discussion that follows, we
assume they know} $v_i$; \update{}{we differentiate cases where they know} $v_i$
\update{}{before
using the data (Section} \ref{par:know}) and after in Section \ref{subsec:nowtp}.

\mypar{Data Model} To define market design we assume a market of
\emph{structured data} and consider other data models later in the paper. In a
market of structured data, $d_i$ are relations and $m$ is a combination of
$d_i$. The arbiter's goal is to identify a function, $F(d_i) = m$, that
transforms a combination of datasets into a mashup. Note the function $F()$ does
not need to use relational operations only. 

\subsection{Market Design}
\label{subsec:marketdesigndef}

A \emph{market design}, $M$, \update{R4D1}{is a collection of 5 components that govern the
interactions between sellers, buyers, and arbiter.}

\mypar{Elicitation Protocol} An elicitation protocol is established between
buyers and arbiter in order to agree on a data transaction. Unlike in
traditional market of goods, the mashup that a buyer desires may not exist
before the arbiter builds it, so this need must be communicated. We introduce a
willing-to-pay function (\textbf{WTP-function}) as a building block of
elicitation protocols. The buyer uses the WTP-function to indicate
the data it needs and how much it is willing to pay for it.


\mypar{Allocation function} At any given time, multiple buyers may want to buy a
particular mashup of interest. The allocation function solves which buyers get
what mashup.

\mypar{Payment function} This function indicates how much money
buyers need to pay to obtain the mashup.

\mypar{Revenue allocation} This function determines how much of the revenue
raised by selling $m$ is allocated to the data that led to $m$. In the case of
markets of relational data, a mashup is a relation, and the revenue allocation
function determines how much of the money raised is allocated to each row in the
mashup.

\mypar{Revenue sharing} Following with the previous example, if $m$ is a
relation, a row in $m$ results from applying $F()$ to the input datasets, $d$.
The revenue sharing function determines how the value allocated to a row in $m$
propagates to $d$. 

\update{R3D2}{The space of market designs is vast. We are interested in market designs that
lead to good outcomes when participants are strategic,} \ie \update{}{they seek to maximize
their own interests. In particular, we are interested in market designs that: i)
guarantee incentive-compatibility; ii) that maximize a market goal, and; iii)
that are practical. We briefly discuss each of the requirements for good market
designs.}


\emph{Incentive-compatible} \update{R3D2}{means that the best strategy the
market players} (\ie
\update{}{buyers and sellers) have is to participate} \emph{truthfully}, \eg
\update{}{it is in
buyers' best interest to declare their true mashup valuation instead of
gaming the market. This property is important for two reasons.
First, it is easy for the players to understand how they participate in the
market, as opposed to having to strategize based on other participants' actions.
Second, it is possible for the market designer} (\ie us) to understand the market
outcomes and therefore make sure it works as we wish.

\emph{Market goals}. \update{R3D2}{A market design can be engineered to maximize
revenue, to optimize social surplus, and others. These goals can be formally
guaranteed. We discuss later how different market environments require different
goals.}

Finally, the market design must be \emph{practical}. This means it can be
implemented in software and serve many participants; \ie the 5 components of a
market design must be computationally efficient.


\subsection{Challenges and Research Agenda}
\label{subsec:mdchallenges}


We discuss next 3 challenges of market designs related to the 5 components we
have described above.

\subsubsection{The Unique Characteristics of Data as an Asset}

\update{R2,R3}{The allocation and payment functions of the market design must
ensure incentive-compatibility from the buyers' side. Data
uniqueness as an asset makes it challenging to design these mechanisms.}

\update{R2,R3,R4}{The allocation and payment functions decide who gets the asset and how much they
pay for it. Auctions offer an example of allocation and payment functions. 
For example, in a generalized second-price
auction}~\cite{optimalauctiontheory}, \update{}{buyers bid for assets and the market
decides who obtains the asset in such a way that the top-K bids are allocated
the K finite assets and each kth-buyer pays the bid made by the (k-1)-buyer.}
Auctions of this kind have been used, among others, to implement real time
ad-bidding that powers today's Internet economy~\cite{adwordsauction}.
The technical details that explain why this mechanism works and elicits truthful
behavior from participants are beyond of the scope of this paper. When
designing mechanisms~\cite{mechanismdesign} to trade data we must pay attention
to its unique characteristics as an asset: data is \textbf{freely replicable}
and it \textbf{can be combined arbitrarily}.

Because data is \emph{freely replicable}, it could be trivially allocated to
anyone who wants it because its supply is infinite. That is at odds with
eliciting truthful behavior from buyers because if buyers know that supply is
infinite they will underbid knowing they will eventually get allocated the
asset. Because data can be \emph{combined arbitrarily} it is difficult to price
a dataset before knowing how it will be used.

Mechanisms to trade digital goods with infinite supply have been proposed
before~\cite{digitalgoods1, digitalgoods2, digitalgoods3}. We are building on
these ideas when engineering market designs.

In particular, we are designing new allocation and payment mechanisms that work
when data is the asset and that are resistant to strategic players who know how
to maximize their utility (the value they get for the data they obtain) over
time. \update{R4D9}{The mechanisms we are designing take into consideration the
externalities generated when trading data. These externalities are related,
among others, to privacy loss by the individual, which is multiplied with the
potential for arbitrary combination of datasets}.


\subsubsection{Elicitation Protocol}

\update{R2D3}{This challenge is concerned with the design of protocols that
allow buyers and arbiter to communicate what data is needed, and how much money
buyers are willing to pay for it. We introduce a WTP-function to achieve this,
and consider two different protocols. The first is appropriate when buyers know
how to value ($v_i$) the dataset a priori, and the second when they only learn
$v_i$ after seeing and using the data.  For example, buyers may sometimes know
the attributes they need to train a classifier or producing a report. Other
times, buyers may need to engage in exploratory tasks before being able to
declare he task they want to solve.}

\paragraph{Buyers know what they want}
\label{par:know}


\update{R4D3}{In its simplest case, a WTP-function contains a superset of the
information required to combine datasets.} In this case, \update{R2W2}{buyers
need to indicate: i) the task they want to solve;} ii) \update{}{a metric to
measure the} \emph{degree of satisfaction} \update{}{that a dataset achieves for
given a task; iii) a price function that indicates how much they are willing to
pay depending on the degree of satisfaction achieved.} For example, a buyer who
wants to train a machine learning classifier can specify the features desired as
attributes of a relation---in a query-by-example type of interface~\cite{dod}.
The buyer can indicate that the metric of interest is accuracy and they may
declare they are willing to pay \$100 for any dataset that permits the model
achieve 80\% accuracy, and \$150 if the accuracy goes beyond 90\%.

\update{R2W2}{Conveying all this information is hard, however, and depends on
the specific task, metric, and pricing strategy}. To model these needs, the 
\textbf{WTP-function} consists of 4 components:

$\bullet$\noindent A package that includes the data task that buyers want to solve. For
example, the code to train an ML classifier. The package is sent to the arbiter, so the arbiter can evaluate
different datasets on the data task and measure the degree of satisfaction.

$\bullet$\noindent A function that assigns a WTP price to each degree of
satisfaction. For example, this function may indicate that the buyer will not pay
any money for classifiers that do not achieve at least 80\% classification
accuracy, and that after reaching 80\% accuracy, the buyer will pay \$100. 

$\bullet$\noindent Packaged data that buyers may already own and do not want to pay money for.
For example, when buyers own multiple features relevant to train the ML model
but want other datasets to augment their data with more features and training
samples, they can send their code and data to the arbiter.

$\bullet$\noindent List of intrinsic dataset properties such as \emph{expiry date} to
indicate for how long data is valuable to them; \emph{freshness} to indicate
that more recent datasets are more valuable; \emph{authorship} to indicate
preferences in who created the dataset; \emph{provenance} to indicate buyer
needs to know how data was generated; and many others such as semantic metadata,
documentation, frequency of change, quality, etc. For example, the buyer may
indicate the need for data not older than 2 months, fearing concept
drift~\cite{conceptshift} will affect classification accuracy otherwise.

\update{R2W2}{There are many interesting challenges around implementing
WTP-functions that fall directly into the realms of database research.}

\emph{Interface} \update{R2W2, R4D3}{We are working on new interfaces to enable
users to easily define WTP-functions. These new interfaces must permit
declaration of data tasks without looking at the data first, for example,
through a schema description}~\cite{dod}, \update{}{ that is similar to
query-by-example interfaces. These new interfaces require new data models to
express not only relational operations but also fusion operations that would
permit merging/contrasting different signals/opinions and transformation needs,
such as pivoting, aggregates, confidence intervals, etc. The WTP-Functions
produced need to be interpreted by the} \emph{mashup builder}, \update{}{a
component we introduce in the next section as part of our DMMS architecture that
is in charge of matching supply and demand.}

\emph{Task Multiplicity} \update{R2W2, R4D3}{Different tasks require different
metrics to measure the} \emph{degree of satisfaction}. \update{}{For example, a
classification task may use} \emph{accuracy} \update{}{as the metric of choice.
A relational query may benefit from notions of} \emph{completeness}
\update{}{borrowed from the approximate query processing
literature}~\cite{aqp1}. \update{}{Dealing with this multiplicity of tasks
introduces interesting research challenges.}

\paragraph{Buyers do not know how much to pay}
\label{subsec:nowtp}

\update{R2,R3,R4}{Sometimes buyers want to engage in exploratory tasks with data without having a
precisely defined question a priori. In these cases, it is not possible for the
buyer to describe the task they are trying to solve. It follows they cannot
express how much they are willing to pay for a dataset.}

\update{}{We are investigating} \emph{ex post} \update{}{mechanisms that work in the following way.
Buyers get the data they want before they pay any money for it. After using the
data and discovering---a posteriori---how much they value the dataset, they pay
the corresponding quantity to the arbiter. In this situation, one may think that
buyers will be motivated to misreport their true value to maximize their
utility. The crucial aspect of the mechanisms we are designing is that they make
reporting the real value the buyer's preferred strategy.}

\subsubsection{Revenue Allocation and Sharing}

The last challenge is concerned with the last two components
of the market design: revenue allocation and revenue sharing. Consider the
arbiter uses a function, $f()$, to create a mashup $m$ from datasets $d_1, d_2$
and $d_3$. The function can be in its simplest case a relational query but it
can also contain non-relational operations. We evaluate the WTP-function
provided by a buyer on $m$ and determine that $m$'s price is $p$ after measuring
the degree of satisfaction. 

\mypar{Revenue allocation} The revenue allocation problem consists of
determining what portion of $p$ is allocated to each row in $m$. Intuitively,
this is asking how valuable is each row in the mashup. Some related work has
modeled this problem as if each row in $m$ was an agent cooperating together
with all other rows to form $m$. Within this framework, the Shapley
value~\cite{shapleyval} has been used to allocate revenue to each row individually
after assuming that the involved datasets participate in a coalition. We are
investigating alternative approaches that are more computationally efficient and
maintain the good properties conferred by the Shapley value.


\mypar{Revenue sharing} The revenue sharing problem determines how the price
from each row in $m$ is shared among the contributing datasets, in the example
above $d_1, d_2$ and $d_3$. The contribution of the datasets to $m$ is
determined by the function $f()$ used by the arbiter in the first place to build
$m$. The revenue sharing problem consists of reverse engineering such function.
Note that if $f()$ is a relational function, then we can leverage the vast
research in provenance~\cite{provenance1, provenance2} to approach the revenue
sharing problem. When $f()$ is a more general function, revenue sharing becomes
challenging. We are investigating information-theory and flow
control techniques to understand how to approach this problem.


\subsection{Market Design Space}
\label{subsec:mdtypes}


The space for market designs is vast and depends on the specific constraints and
goals that we want the market to honor. We consider below different markets and
discuss how they motivate different market designs. 

\mypar{External markets} In external markets, independent organizations trade
data assets. In this case, money is a good incentive to
get companies who own valuable information to share it with others that may
benefit from its use. One possible market design for this scenario is to
maximize revenue. The arbiter extracts as much money from buyers as possible so
it can use that money to incentivize sellers to share their data. Variations of
this market may allow sellers to set a reserve price; sellers will not sell any
data unless they obtain a given quantity. Achieving these goals requires
careful design of the 5 components of a market design.

\mypar{Internal markets} In internal markets, members of an organization share
data internally to maximize data's value. Internal data market platforms have
the promise of bringing down data silos by incentivizing data owners (\eg
specific teams, or individuals) to publish their data in a way that is easy to
consume by others, in exchange for some reward. \update{R4D4}{In this scenario,
it is reasonable that a market design optimizes social welfare, that is, the
allocation of data to buyers. The mechanisms to incentivize sellers
can take the form of bonus points, or time.}

\mypar{Barter and Gift Markets} These are markets where the participant's
incentive to share their data is to receive data or services from somebody else.
\update{R4D6b}{Existing barter markets are coalitions of
hospitals}~\cite{hospital1, hospital2}, \update{}{but also the exchange of
personal data for Internet services, such as social media platforms, etc. Market
designs for this kind of market will differ from external and internal
markets.}

\update{R4D1}{The variety of market designs one may want to deploy calls for
plug'n'play interfaces. We aim to design data-market management systems (DMMS)
that permit the declaration of a wide variety of market designs to cater for
different scenarios and their deployment on the same software platform (see}
\F~\ref{fig:marketdesigntimeline}). We explain the architecture of the DMMS in
the next section. Before that, we include a FAQ and a summary of this
section.

\subsection{FAQ: Frequently Asked Questions}

\noindent\textbf{Why would people use the market to share data?} A well-designed
market incentivizes sellers to share data to obtain some profit, which may be
monetary or some other form. It also incentivizes buyers to share their data needs in
exchange for having their discovery and integration problems solved by the
arbiter.

\noindent\textbf{What if I am not sure if my dataset is leaking personal
information?} Sharing data is predicated on the assumption that it is legal.
Certain PII information, for example, cannot be shared across entities without
users' permission. The DMMS that we present in the next section offers tools
that help to reduce the risk of leaking data.

In addition, once a dataset has been assigned a price, it is possible to
envision a data insurance market, where a different entity than the seller
(i.e., the arbiter) takes liability for any legal problems caused by that data.
In this case, the arbiter is incentivized to avoid those problems, stimulating
more research in secure and responsible sharing of data.

\noindent\textbf{Wouldn't markets concentrate data around a few organizations
even more?} Today, data is mostly concentrated around a handful of companies
with the expertise and resources to generate, process and use it. Ideally, we
want to design markets that bring the value of data to a broader audience. It
is certainly possible that a market would only worsen this concentration by
allocating data to the richest and more powerful players. Fortunately, it is
possible to design markets that disincentivize this outcome: achieving that is a
goal of our research.

\noindent\textbf{Is there going to be enough demand for a given, single dataset?}
We expect certain datasets will naturally have less demand than others, as with
any asset today. However, with a powerful enough arbiter, individual datasets
are combined and add value to lots of different mashups that may be, in turn,
designed to satisfy a varied set of buyers' needs. 

Furthermore, studying the market dynamics will be important to determine, for
example, if domain-specific markets (markets for finance, for health, for
agriculture) would be more efficient than more general ones in concentrating and
uncovering highly valuable datasets.

\noindent\textbf{Why would a seller or buyer trust the arbiter?} We do not assume
they would, and we discuss in the next section how this is a key design goal of
a DMMS.

%

\noindent\textbf{The arbiter could prevent data duplication by assessing what
datasets to accept, hence addressing one of the challenges of selling data.}
Regardless of the merits of that mechanism to enforce the right outcomes in the
market, this design would not allow participants to trade freely. Furthermore,
since datasets can be arbitrarily similar to each other, it is unclear what
threshold the arbiter should use to make a decision, or how to compute that
threshold in the first place.

\noindent\textbf{Why would a buyer give out their code (as part of the
WTP-function) when it may be an industrial secret?} 
\update{R2D3}{If a buyer knows how to specify a WTP-function for their task and they are
willing to give the code away, then the mechanism explained above works. If the
buyers do not know how to specify a WTP-function or they are not willing to give
their code away, they can use the mechanism we discuss in
Section}~\ref{subsec:nowtp} \update{}{to obtain the data from the arbiter and run the code
locally. The challenge in this situation is related to designing }\emph{truthful}
\update{}{mechanisms that incentivize buyers to tell the real value instead of reporting a
low value to maximize their utility.}



%
%

\section{Data Market Managmt. System}
\label{sec:platform}

\begin{figure}[t]
  \centering
  \includegraphics[width=\columnwidth]{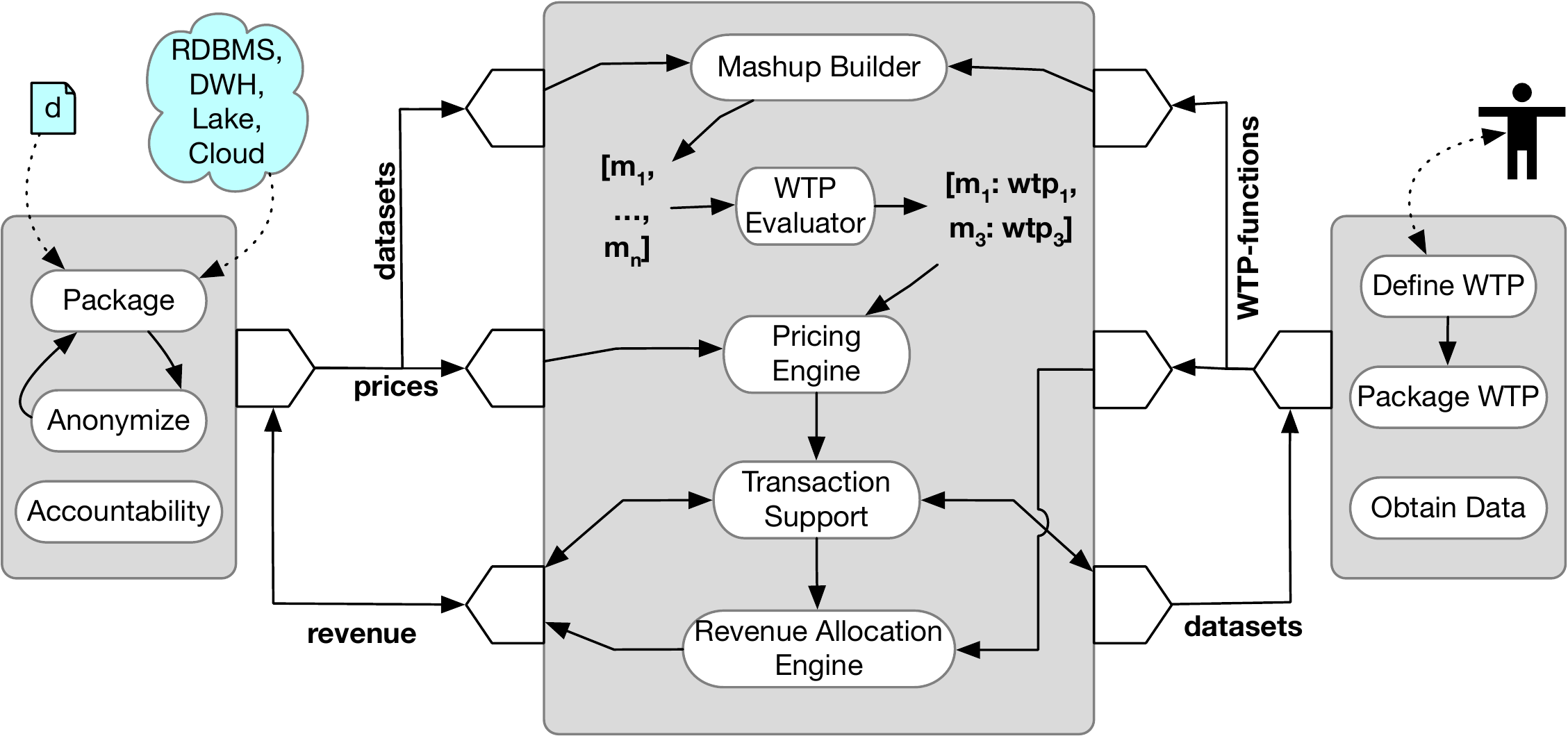}
\caption{Architecture of a Data Market Management System.}
\label{fig:architecture}
\end{figure}

Data market management systems must be designed to support different market
designs and they must offer software support to sellers, buyers, and the
arbiter. The DMMS system we propose achieves that using a seller, buyer, and
arbiter management platforms, which are shown in \F\ref{fig:architecture}.

\subsection{Overview of Arbiter Managmt. Platform}
\label{subsec:overviewams}

The arbiter management platform (AMP) is the most complex of all DMMS's
components: it builds mashups to match supply and demand, and it 
implements the five market design components. We use the architecture in
\F\ref{fig:architecture} to drive the description of how the AMS works.

The AMS receives a collection of WTP-functions from buyers specifying the data
needs they have. Sellers share their datasets with the arbiter, expecting to
profit from transactions that include their datasets. The AMS uses the
\textbf{Mashup Builder} (top of the figure) to identify combinations of
datasets (\ie \emph{mashups}) that satisfy buyers' needs. These are
depicted as $[m_1, m_2, ..., m_n]$ in the figure. 

The next step is to evaluate the degree of satisfaction that each mashup
achieves for each buyer's WTP-function. This task is conducted by the
\textbf{WTP-Evaluator}. The WTP-Evaluator first runs the WTP-function code on
each mashup and measures the degree of satisfaction achieved. With the degree of
satisfaction, it then computes the amount of money (or other incentives) the
buyer is willing to pay, $wtp_i$. The output of the WTP-Evaluator is a
collection of pairs $m_i, wtp_i$ indicating the amount of money that a buyer is
willing to pay for each mashup that fits the needs indicated by their
WTP-function.

The next step is to use the \textbf{Pricing Engine} to set a price for each
$m_i$ and choose a winner\footnote{the market design may specify more than
one winner, but we use one here to simplify the presentation}. The
\textbf{Transaction Support} component delivers $m_i$ to the winning
buyer and obtains the money, $wtp_i$. Finally, the \textbf{Revenue Allocation
Engine} allocates $wtp_i$ among the sellers that contributed datasets used
to build $m_i$ and the arbiter. At this point the transaction is completed.


\mypar{Arbiter Services} Because the arbiter knows the supply and demand for
datasets, it can use this information to offer additional services for buyers
and sellers, perhaps for a fee. For example, the arbiter could recommend
datasets to buyers based on what similar buyers have purchased
before~\cite{collaborativefiltering}. \update{R4D1}{This kind of service,
however, leaks information that was previously private to other buyers. This
information leak affects the elicitation, allocation, and payment functions of
the market design (see Section}~\ref{subsec:marketdesigndef}). \update{}{To see
why, consider how a buyer may change their bidding strategy if they knew 
the price at which a dataset was sold recently.}

\mypar{Negotiation Rounds} If the AMS cannot find mashups that fulfill the buyer's
needs, it can describe the information it lacks and ask the sellers to complete
it. Sellers are incentivized to add that information to receive a profit. For example,
the AMS may ask the seller to explain how to transform an attribute so it joins
with another one, or it may request information about how a dataset was
obtained/measured, semantic annotations, mapping tables, etc. Sellers will be
incentivized to help if that raises their prospect of profiting from the
transaction. Similarly, buyers can request the arbiter for data context
(provenance, how data was measured/sampled, how fresh it is, etc.) when they
need it to use the data effectively. 

\subsection{Seller Management Platform}

The SMP communicates with the AMS to share datasets and receive profit, to
coordinate private data release procedures (as we see next), as well as to agree on
changes to the dataset that may improve the seller's chances of participating in
a profitable transaction. Next, we explain the key services we envision SMP
offering sellers:

\mypar{Statistical Database Privacy} Even if incentivized to sell data for
money, sellers face a deterrent when their data may leak information---\eg
personally identifiable information (PII)---that should not be public.  To
assist sellers, the SMP must incorporate some support for the safe release of
such sensitive datasets. And because datasets may leak information
when combined with other datasets~\cite{deanonymizenetflix}---which is precisely
what the arbiter will do as part of the mashup building process---the protection
process must be coordinated between SMP and AMS. \update{R2W1}{This component
introduces opportunities to leverage the rich literature on differential
privacy}~\cite{privacypreservingdata} and variants~\cite{pufferfish}.
\update{}{Applying statistical database privacy will reduce data's value for
buyers who want to access precisely those records. It is an open area of
research to understand how to price the externality that the loss in privacy
introduces, a question other work has started exploring}~\cite{ppd14}. The
challenge is on identifying a good balance between protection and profit.


\mypar{Accountability} The SMP must allow sellers to track how their datasets
are being sold in the market, \eg as part of what mashups. \update{R2W3}{When
sellers update datasets in the platform, the SMP incrementally updates the
information recorded about those datasets subject to an optional access quota
established by the origin system. This permits the SMP to maintain fine-grained
lineage information that is made available on demand.}

\mypar{Data Packaging} The SMP transforms datasets provided by sellers into a
format interpretable by the arbiter. In addition, this feature allow sellers to
share datasets in bulk by pointing to a data lake, cloud storage full of files,
a databases, or data warehouse. This functionality is useful in internal data
markets to unlock data silos.

\subsection{Buyer Management Platform}

Data buyers must provide the arbiter with a willing-to-pay function
(WTP-function) that indicates the price a buyer is willing to pay given the
satisfaction achieved by a given dataset. Buyer management platforms (BMP) have
the following requirements:

\begin{myitemize}
\item Because manually describing a WTP function may be difficult, a BMP must
help buyers define it through interfaces that permit descriptions of a
multiplicity of tasks (see Section~\ref{par:know}). 
\item Secure sharing of the WTP function with the arbiter, so the arbiter 
computes the level of satisfaction of different mashups and obtains the WTP price
buyer bids for such a mashup. 
\item Finally, a communication channel enables buyer-arbiter exchange mashups,
WTP-functions, as well as allow the arbiter to recommend alternative datasets to
the buyer, \eg when the arbiter knows of other similar buyers
who have acquired such datasets.
\end{myitemize}

\subsection{Trust, Licensing, Transparency}

Now we zoom out to the general architecture comprising AMS, BMS, and SMS and
consider how differing degrees of trust, the existence of data licenses, as well
as the need for transparency, introduce additional challenges for the design and
implementation of a DMMS.

\mypar{Trust} \update{R3D1}{We have assumed so far that sellers and buyers trust the arbiter.
Sellers trust that the arbiter will not share the data without sellers' consent,
that it will implement the rules established by the market design faithfully,
and that it will allocate revenue following those rules too. Buyers trust the
arbiter with their code (that ships as part of the WTP-function), and similar to
sellers, they trust the arbiter will enforce the agreed market rules.
We think this trust is generally granted in the context of
internal markets. In the context of external markets, although we think it is
reasonable to assume trust in a third party---similar to how individuals and
organizations trust the stock market---it is conceivable to imagine scenarios
where trust is not granted. In this case, we need to consider techniques on
privacy-preserving data management}~\cite{privacypreservingdata}, processing
over encrypted data~\cite{processingencrypteddata}, as well as decentralized,
peer-to-peer markets and blockchain platforms~\cite{blockchain1, blockchain2}.
\update{R3D1}{Using the above techniques it is conceivable to engineer the
arbiter so it is not a logically centralized entity anymore, hence aiding buyers
and sellers to gain trust and participate in the market.}
\update{R2W1}{Introducing these techniques protects data owners at the cost of
reducing data's value. Others have explored this }tradeoff~\cite{chenMLQ19,
revMax19, ppd14} \update{}{making an explicit connection between privacy and data
value.}
Finally, for situations when the chain of trust is broken, dispute
management systems must be either embedded in or informed by the transactions
that take place in the DMMS so the appropriate entities can intervene and
resolve the situation.

\mypar{Data licensing} Sellers can assign different licenses to the datasets
they share that would confer different rights to the beneficiary. Similarly,
buyers may be interested in obtaining datasets subject to licensing constraints.
For example, a hedge fund may want to acquire a dataset with
exclusive access, preventing perhaps other competitors to access the same data.
The artificial scarcity generated by this license should cost more to buyers,
who could be forced to pay a 'tax' so long they maintain the exclusivity access.
Other types of licenses are those that transfer ownership completely, so buyers
sell the datasets as soon as they have bought them (creating a market for
arbitrageurs as we discuss in the next section), or licenses that prevent the
beneficiaries from selling a previously acquired dataset. Supporting these
licensing options affects both market design and DMMS system.  Furthermore, it
raises questions of legality and ethics that go beyond computer science and
economics. Concretely, we are exploring software implementations of contextual
integrity~\cite{contextualintegrity}, which we believe may be an interesting
vehicle to enable data licensing.

\mypar{Transparency} Transparency may be required at many points of the market
process. Sellers may need to know in what mashups their data is being sold and
what aspects of their data (rows, columns, specific values) is more valuable.
Similarly, buyers may request transparent access to the mashup building process
to understand the original datasets that contribute to the mashup and decide
whether to trust them or not. We do not discuss the implications of these
requirements, we only highlight they have an impact on the engineering of a
DMMS.

\subsection{Markets of Many Data Types}

We have presented the AMP, SMP, and BMP without focusing on a specific type of
data to be exchanged. We envision markets to trade data of many types:

\mypar{Multimedia Data} A variety of multimedia data such as text, web (\ie a
search engine market that does not depend on ads?), as well as images and video
are likely targets for a data market platform. How to build DMMS platforms to
reason about how to combine and prepare this data for buyers is a challenge.

\mypar{Markets for Personal Data} Ultimately, we would like to be able to price a
person's own information. If I knew how much the information I am giving an
online service is worth, I could make a better decision on whether the exchange
is really worth it or not. Because many times an individual's own data is not
worth much in itself---but quickly raises its value when aggregated with other
users---it is conceivable that coalitions of users would form who collectively
would choose to relinquish/sell certain personal information to benefit together
from their services. Some are advocating for these so-called \emph{data
trusts}~\cite{datatrust1, datatrust2}.

\mypar{Embeddings and ML Models} Embeddings and vector data are growing fast
because they are the input and output format of many ML pipelines. As
data-driven companies keep building on their ML capabilities, we expect this data
will only grow. Obtaining some of these embeddings incurs a high cost in
compute resources, carbon footprint, and time. For example, the BERT pre-trained
models produced by Google~\cite{bert} take many compute hours to build. For this
reason, we expect companies will rely on the exchange of pre-trained embeddings more
and more, and hence our interest in supporting this format in our data market
platforms. More generally, this motivates markets for data products---\ie software,
data, and services derived from data.

We focus initially on tabular data such as relations and spreadsheets because
this data is sufficient to cover most business reporting, analytical, as well as
many machine learning tasks. In the next section, we introduce a \textbf{Mashup
Builder} specific to this type of data.

\section{MB: Matching Supply and Demand}
\label{sec:mashup}

The goal of the mashup builder is to generate a collection of mashups that
satisfy a WTP-function. \update{R4D5}{Key research goals of our vision are to
understand the extent to which this process can be fully automated, and devise
the best ways of involving humans when it is not possible (see}
Section~\ref{subsec:machinesandpeople}). The architecture of the system we are
building is depicted in \F\ref{fig:architecturemashup} and is designed to
address the following problems:

\begin{figure}[t]
  \centering
  \includegraphics[width=\columnwidth]{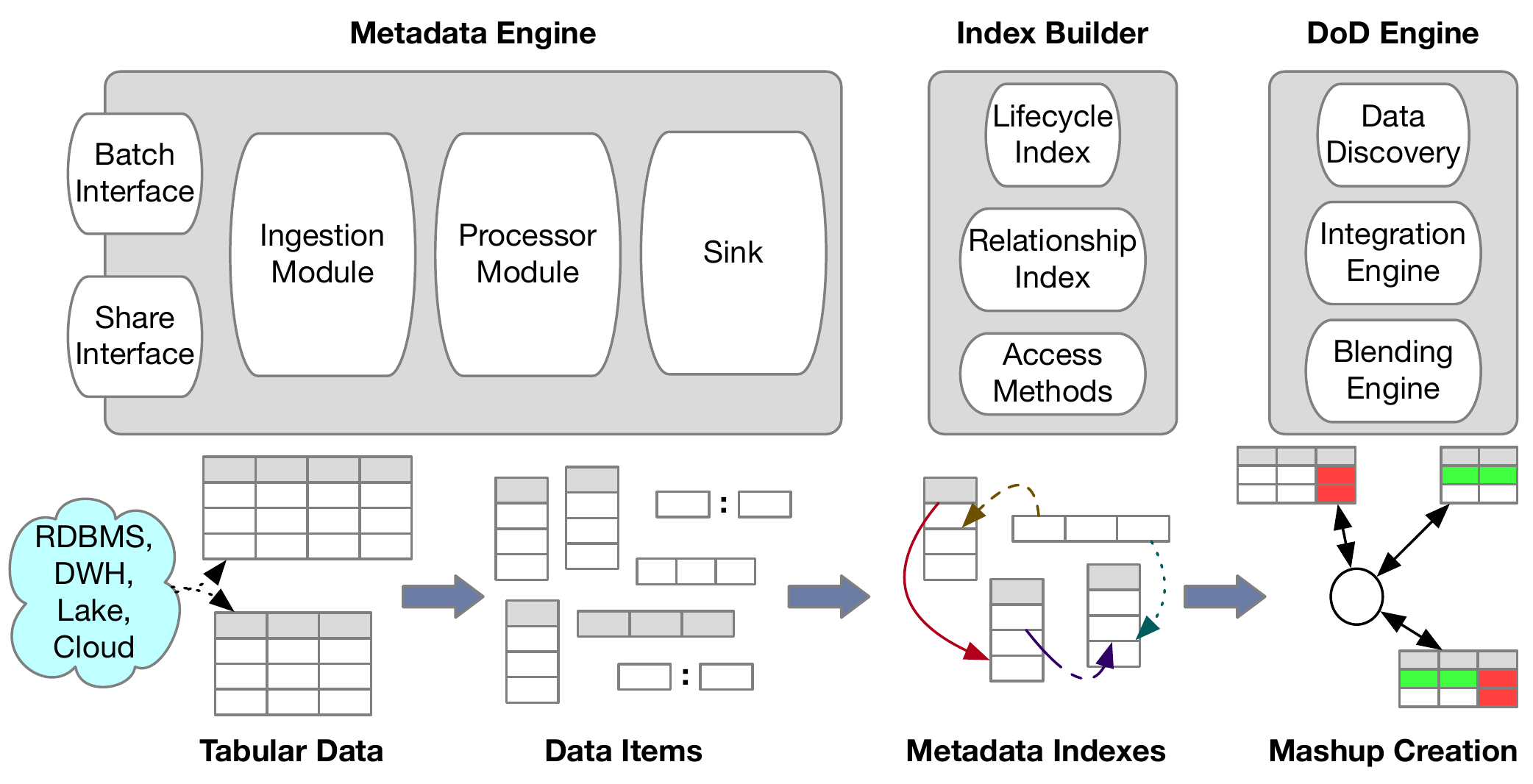}
\caption{Architecture of Mashup Builder: metadata engine, index builder and DoD
engine.}
\label{fig:architecturemashup}
\end{figure}

\smallskip

\mypar{Data Discovery} The arbiter receives datasets from sellers, some of whom
may be organizations with thousands of datasets. The goal of data discovery is
to identify a \emph{few} datasets that are relevant to a WTP-function among
thousands of diverse heterogeneous datasets.

\smallskip

\mypar{Data Integration and Blending} The goal of data integration and blending
is to identify strategies to combine the datasets identified by the discovery
component into mashups that satisfy the WTP-function. 
Those strategies consist of identifying mapping and transformation functions to
join attributes as well as other preparation tasks such as value interpolation
to join on different time granularities.

Because multiple similar datasets may contribute to the same or a small group of
similar mashups, \textbf{data fusion} operations permit combining and
contrasting the different combinations, keeping track of the origin of each data
item, so consumers understand how data was assembled. 

\smallskip

We bootstrap the implementation of the mashup builder with Aurum~\cite{aurum}, a
data discovery system that allows users to find relevant datasets and to combine
them using join operations. To do that it extracts metadata from
the input datasets, it organizes that metadata in an index and uses the index to
identify datasets based on the criteria indicated in the WTP-function.  The
architecture of the Mashup Builder is shown in \F\ref{fig:architecturemashup}.
We describe the components next:

\subsection{Metadata Engine}

The metadata engine's goal is to read and maintain the lifecycle of each input
dataset. Datasets can be automatically read from a \emph{source} in bulk (\eg a
relational database, a data lake, a repository of CSV files in the cloud) or
they can be registered manually by a user who wants to share specific datasets.
\update{R2W3}{When datasets change in the sources,} \eg \update{}{when users manually
update them, the metadata engine registers and keeps track of these changes,
maintaining version history}. All these tasks are performed by the
\textbf{ingestion module} through its batch and sharing interfaces as shown in
the figure. Each dataset is divided conceptually into \emph{data items}, which
are the granularity of analysis of the engine. For example, a column data item
can be used to extract the value distribution of that attribute. A row data item
can be used to compute co-occurrences among values. A partial row data item can
be used to compute correlations, among others. For each dataset, the metadata
engine maintains a time-ordered list of context snapshots. \update{R2W3}{A
context snapshot captures the properties of each dataset's data item
at each point in time}. For example, signatures of its contents, a collection of
human or machine owners (\ie what code is using what data), as well as the
security credentials. This is performed by the \textbf{Processor} component of
the system.


Because data item information is not given directly at ingestion time, the
engine must harness that information. Data market platforms aim to incentivize users to
provide that information directly, but in certain scenarios this is not
possible: \eg a data steward pointing to a collection of databases in an
internal organization. The output of the metadata engine is conceptually
represented in a relational schema managed by the \textbf{Sink} component.

The metadata engine is a fully-incremental, always-on system that maintains
the output schema as updated as possible while controlling the
overhead incurred in the source systems and the precision of the output
information.


\subsection{Index Builder}

The index builder processes the output schema produced by the metadata engine
and shapes data so it can be consumed by the dataset-on-demand engine (DoD),
which is the component in charge of integration and blending of mashups.
Among other tasks, the index builder materializes join paths between files,
and it identifies candidate functions to map attributes to each other; \ie it
facilitates the DoD's job.  The index builder keeps indexes up-to-date as the
output schema changes. This calls for efficient methods to leverage the
signatures computed during the first stage.

\subsection{DoD Engine}

The DoD engine takes WTP-functions as input and produces mashups that fulfill
the WTP-function requests as output. It uses the indexes built by the index
builder, the output schema generated by the metadata engine, as well as the
raw data.

\smallskip

The DoD relies on query reverse engineering and query-by-example techniques
\cite{reversequery1}, as well as program-synthesis~\cite{programsynthesis},
among others, to produce the desired mashups.

\smallskip

\mypar{Data Fusion} When there are many datasets available, DoD may find
multiple alternatives to produce mashups. In certain cases, a buyer wants to see
a contrast of mashups (this will be specified in the WTP-function). For example,
consider a buyer who wants to access weather data and there are multiple sources
that provide this information. A data fusion operator can align the differing
values into a mashup that the buyer can explore manually. A specific fusion
operator may select one value based on majority voting, for example, while other
fusion operators will implement other strategies. Buyers may want to have access
to all available signals to make up their own minds. As a consequence, buyers may
want to use DoD's fusion operators to help combine the different sources into
mashups. 


\subsection{Machines and People}
\label{subsec:machinesandpeople}

\update{R4D5}{Automatically assembling a mashup from individual datasets when only given a
description of how the mashup should look is an ambitious goal. Our
experience working on this problem for the last few years has taught us that in
certain cases this may not be possible at all, such as when ambiguity makes it
impossible to understand the right strategy to combine two datasets. We devise
two strategies to tackle this problem.}

\mypar{Request additional input when automatic integration does not work} The
first strategy is to have the AMS system interact with sellers to request
additional information about the datasets they have shared that may help with
the integration and blending process, \eg a semantic annotation, a function to
obtain an alternative representation, etc.  Sellers willing to include the
additional information can be incentivized to do so by obtaining a higher
profit.

\mypar{Involve humans in the loop} Another strategy is to directly
incorporate humans-in-the-loop as part of the mashup builder's normal
operation. This has been done to answer relational queries~\cite{crowddb,
crowdenumeration}, and there are opportunities to extend those techniques to
help with integration and blending operations as well. Because all this takes
place in the context of a market, it becomes possible to compensate humans
according to the value they are creating.

\section{Evaluation Plan}
\label{sec:evaluation}

In this section, we explain how we plan to evaluate market designs, as well as
the DMMS implementation.

\subsection{Simulation of Market Designs}
\label{subsec:simulation}

A market design that is sound on paper may suffer unexpected setbacks in
practice. This may happen because rationality assumptions made at design time
may break in the wild. In the context of mechanism design/game theory,
rationality is interpreted as players will play the best strategy available to
them.  Unfortunately, that does not account for risk-lover or ignorant players.
Furthermore, some players may be adversarial in practice, forming coalitions
with other players to game the market. Or less dramatic, a faulty piece of
software may cause erratic behavior. Below, we explain how we plan to evaluate
the effectiveness and efficiency of market design in practice.

\mypar{Effectiveness} The mismatch between theory and practice calls for a
framework to evaluate how resilient a market design is under adversarial, evil,
and faulty processes. We plan to design a simulation platform where it is
possible to implement different rules and change the behavior of players, and
where it is possible to model adversarial, coalition-building, as well as risky
and ignorant players (this is shown in (3) of \F\ref{fig:marketdesigntimeline}).
The simulation platform will test a market design's robustness before
deployment.

Large-scale simulations introduce database challenges such as: i) supporting
quick communication among many players (transaction processing); ii) modeling
workloads to simulate different strategy distributions of players. Such a
simulation framework will be of independent interest.

\mypar{Efficiency} Market mechanisms are implemented with an
algorithm. The fields of mechanism design and algorithmic game theory have
contributed efficient approximation algorithms~\cite{mechanismdesign}. In
databases, algorithms with high complexity are often used in
practice for small problems, and conversely, sometimes algorithms with low complexity
cannot be used practically because of the data size. We want to contribute
empirical evaluations of these designs.

\subsection{Building and Evaluating a DMMS} 
\label{subsec:builddmms}

\update{R4D7}{The vision is expansive and touches upon different disciplines. We think the
database community is in a great position to refine, correct, expand, and
contribute ideas towards our vision, hence the motivation for this vision paper.
We layout next our research plan:}

\mypar{Market design} \update{}{Because of the unique characteristics of data and the
nature of the platform we propose, we need new theoretical designs that work in
practice. We are in the process of engineering market designs for data.}

\mypar{Market platform} \update{}{We consider the mashup builder a key component of our
vision. We have experience building data discovery and integration systems and
are using this experience to build this next generation of systems.}

\mypar{Market simulation} \update{}{As explained above, we must have mechanisms to
understand the properties of the markets we design before implementing them in a
platform.}

\update{}{Among the multiple ways of creating a DMMS our choice is to start with internal
markets. This will help us hone the interface with humans, understand the
deployment context and its constraints better (e.g., issues of privacy, trust,
licensing), as well as to conduct qualitative evaluations.}

%
%
%

\section{Societal Impact of Data Markets}
\label{sec:impact}

The side effects of data markets span beyond computer science and economics.  We
plan to engage with the broader community of scholars at The University of
Chicago and elsewhere to discuss and outline the challenges of data markets in a
broader societal context. We outline some interesting aspects below.

\subsection{Economic Opportunities}
\label{subsec:opportunities}

A well-functioning market generates economic opportunities for other players
besides sellers and buyers:

\mypar{Arbitrageurs} They play seller and buyer at the same time.  Arbitrageurs
buy certain datasets, transform them, perhaps combining them with certain
information they possess, and sell them again to the market. The transaction
generates a profit for them whenever the sold dataset is priced higher than the
dataset they buy. Since we want to design mechanisms that price datasets based
on supply and demand, it is conceivable that the participation of arbitrageurs
in the market will rise data's value, because they will be incentivized to
transform datasets into a shape that is desired by buyers.

\mypar{Opportunistic data seller} Opportunistic data sellers may not own data,
but they have time that they are willing to invest in collecting high-demand
datasets. They obtain information about highly demanded datasets from the
arbiter.  For example, consider one more time the example of the introduction
with the two sellers and the buyer. Consider a third seller, Seller 3, who does
not own any dataset, but has time, and is willing to use that time to
acquire/find data for profit. Because the arbiter knows that $b_1$ would benefit
from attribute $\langle e \rangle$, which neither $s_1$ nor $s_2$ contain, the arbiter can ask
Seller 3 to obtain a dataset $s_3 = \langle e \rangle$ for money. Because the arbiter knows
supply and demand, not only does it help sellers and buyers, but it creates an
ecosystem of economic opportunities for other entities.  

\mypar{Offloading tasks} As discussed above, when the arbiter does not know how
to automatically assemble a mashup, it can schedule humans to help with the task
and compensates them appropriately for their labor.

\mypar{Data Insurance} Once data has a value and a price, it is possible to
build an insurance market around it. Such an insurance market would be useful to
reason about data breaches, for example. How liable is a company that suffers a
data breach that results in leaking private customer information?  Or, if a
seller shares a dataset that is later de-anonymized by a third party, despite
the best efforts from the arbiter to protect it, who is liable?  Can/Should
insurance cover these cases?

\subsection{Legal and Ethical Dimension}

Who owns a dataset? Throughout this paper, we have assumed that sellers owned
the data they were sharing with the arbiter. Consider a seller who has collected
a dataset through their manual effort and skill. In this case, does the seller
own such a dataset? What if the records in the dataset correspond to users
interacting with a service the seller has created? Do those users own part of
the data too?  A recent article from the New York Times~\cite{nyt-privacy} has
illustrated in glaring detail how it is possible to determine with high
precision the location of individuals and their daily activities from smartphone
data traces. The data that permits that is routinely collected and sold by
companies that profit from it. This leads to questions around what data is legal
to possess, what does it mean to own data, and when it should be possible to
trade data.

\mypar{Market Failures} Markets sometimes fail and cause social havoc. Other
times, markets work only for a few, causing or accentuating existing inequality.
All markets are susceptible to these kinds of problems, including the ones we
envision in this paper. The difference is that we haven't implemented our market
yet, so we have a chance to study beforehand what the consequences of
malfunctioning markets on society are and decide whether the tradeoffs are worth
it. Forecasting the implications of different market designs is a key aspect of
our vision; hence the simulation framework introduced in the previous section.

\section{Related Work}
\label{sec:relatedwork}

We propose the first comprehensive vision of end-to-end data market platforms
that makes an explicit separation between design and implementation (DMMS).  We
start this section with a discussion of data markets
(Section~\ref{subsec:whole}) and then focus on work related to the DMMS and the
Mashup Builder in Section~\ref{subsec:software}.

\subsection{Markets of Data in the Wild}
\label{subsec:whole}

\mypar{Existing Marketplaces of data} We use Dawex~\cite{dawex} as an
illustrative example of platforms that refer to themselves as data markets.
Others include OnAudience.com~\cite{ads1}, BIG. Exchange~\cite{ads2},
BuySellAds~\cite{ads3} for ad data, as well Qlik Datamarket~\cite{mkt1},
Xignite~\cite{mkt2}, WorldQuant~\cite{mkt3}, DataBroker DAO~\cite{mkt4},
Snowflake's Data Exchange~\cite{snowflake-exchange}, among others. In Dawex,
sellers offer datasets that buyers can obtain for a fixed price after seeing a
sample of the data. Dawex acts as a sharing platform but does not solve the
discovery, integration, or pricing problems. Buyers need to commit to pay a
price before truly knowing the value of the dataset. These characteristics are
typical of today's data markets.


\smallskip
\mypar{Data is traded today, everywhere} \update{R4D7}{Data trade is pervasive.
Consider the barter exchange in which users of platforms such as Facebook,
Twitter, Google, or many others take part. Users give away their data in
exchange of the service provided by the companies. Health institutions such as
hospitals routinely share data to improve patient care and
treatment}~\cite{hospital1, hospital2}.  Consider the complex economy of
data brokers~\cite{databroker1, databroker2} \update{}{who, albeit not directly
connected to everyday consumers, trade data with the large corporations to
profit. In many of these cases, the users who contribute their data are not
aware of how it is being used behind the scenes}~\cite{inetservice1,
inetservice2}.
Recently, the concept of a data trust~\cite{datatrust1} \update{}{has been proposed to
treat data as labor and let users gain control of their data by pooling it
together with other users. The data markets we propose make data trading
explicit, and measuring data's externality is a component of our vision. We
think these two characteristics bring transparency to an otherwise opaque
economy.}


\subsection{Theoretical Data Markets}
\label{subsec:theorymarkets}

We discuss theoretical designs related to the components of a market design we
discussed in Section~\ref{sec:design}.

\smallskip
\mypar{Allocation and Payment functions} The marketplace for data
proposal~\cite{agarwal19} models a market that solves the the allocation and
payment problem in a static scenario with 1 buyer and multiple sellers. In
\cite{moor}, the authors propose an end-to-end market design that considers
buyers and sellers arriving in a streaming fashion. This has been an active area
of research modeled by work in dynamic mechanism
design~\cite{dynamicmechanismdesign}, revenue
management~\cite{revenuemanagement}, and algorithmic mechanism design. In the
latter, the work on online auctions~\cite{onlineauctions1, onlineauctions2,
onlineauctions3} for digital goods started an interesting line of research that
has contributed results to ad auctions, among others. We are building on top of
this rich literature with an emphasis for the trade of data when players are
strategic over time and with the ability to construct mashups, a key component
to avoid thin markets, where insufficient number of participants make trade
inefficient. 


\smallskip
\mypar{Revenue allocation} A recent line of work has modeled certain machine
learning settings as coalitions of players contributing (training) data to a
model~\cite{ghorbani2019data, jiaNN19, agarwal19}. To assign credit to the input
datasets, they use the Shapley value~\cite{shapleyval}. Other work suggests 
using a different metric, the core~\cite{core} which is also apt for coalitional
games. Due to the complexity of computing the Shapley value, the contributions
of these papers are usually approximations with good performance guarantees.

\smallskip
\mypar{Query Pricing} There is a long and principled line of work coming from
the database community around the problem of how to price queries
\cite{chenMLQ19, revMax19, qbdp15, ppd14}. In this setting, a dataset has a set
price. The problem is how to price relational queries on that dataset in such a
way that arbitrage opportunities (obtaining the same data through a different
and cheaper combination of queries) are not possible. Recent work in this line~\cite{revMax19} also considers how to
maximize revenue for the broker under the same pricing model as above. If all
datasets of a market are thought of as views over a single relation, then the
setting of this work resembles ours. However, many data integration tasks
require arbitrary data transformations, and many buyers want to buy fused
datasets that contain diverging opinions, for example. This line of
work is complementary to our vision and we plan to include these ideas as part
of our design.

\smallskip
\mypar{Value of Data} An increasing amount of work from the economics literature
focuses on understanding the value of data for a particular
firm~\cite{bigdatafirm18, predwbig13, ebaytranslate, varianio}, an entire
economy~\cite{macro1, macro2}. This work is complementary to ours and relevant
to understand, among others, how an individual or firm identifies their private
valuation. 


\mypar{Privacy-Value Connection} \update{R2W1}{This line of work makes a connection between
data value and privacy} \cite{chenMLQ19, revMax19, ppd14}. \update{}{The buyer can specify
a level of privacy associated with a query, in such a way that the higher the
privacy level, the less the dataset is perturbed, meaning the dataset will be of
higher quality. Therefore, the higher the privacy level, the higher the price of
the dataset.}

In our vision we want to directly link market design with software platforms
(DMMS) to provide an end-to-end market environment with rules governing every
aspect and participant.
We are interested in the engineering of plug'n'play platforms that
can accept different market designs tailored to different scenarios, \eg
internal vs external markets. Although we expect to benefit from many of the past
ideas, the database community faces unique challenges to build practical
data markets.

\subsection{DMMS Related Work}
\label{subsec:software}

We discuss theory, algorithms, and systems for discovery and integration that
are related to the Mashup Builder.


\smallskip
\mypar{Data Sharing Platforms} The datahub system~\cite{datahub} introduced a
data version control system implemented on a software platform that allows
members of a team to collaborate. OrpheusDB~\cite{orpheus} similarly offers
teams the ability to collaborate over a relational system and capture how data
evolves. Outside the database community, there are many sharing systems such as 
TIND~\cite{tind}, KOHA~\cite{koha}, as well as online repositories such as the
Harvard Dataverse~\cite{dataverse} or the ICPSR~\cite{icpsr} at the University
of Michigan, geared towards sharing data across the social sciences. Within
organizations, data warehouses and lakes~\cite{datalake1, datalake2} play the
role of central hubs that facilitate data sharing. 

\mypar{AnyLog} \update{R4D8}{The AnyLog system}~\cite{anylog} \update{}{proposes a platform to pool
Internet-of-Things data into a logically decentralized repository. The paper
emphasizes two key design decisions that are aligned with our data markets
vision. First, participants are incentivized to share their data in exchange for
rewards. Second, market forces guide data owners to curate their data in a way
that is desired by data consumers. Dealing with the integration problem at the
source is simpler than downstream. This idea is related to the} \emph{Negotiation
Rounds} \update{}{between arbiter and sellers we described in
Section}~\ref{subsec:overviewams}. \update{}{We believe many of the ideas we proposed in
this vision paper will complement efforts such as AnyLog and the DMMS we are
building.}

\mypar{Data Discovery} Data discovery systems such as
Infogather~\cite{infogather}, Google Goods~\cite{goods} and Dataset
search~\cite{datasetsearch}, define a specific task and focus on how to build
indexes to solve that task. There is also a line of work on data catalogs, with
Amundsen~\cite{amundsen}, WhereHows~\cite{wherehows}, Databook~\cite{databook}
as open-source examples and Alation~\cite{alation}, Azure's data
catalog~\cite{azuredatacatalog}, and Informatica's data
catalog~\cite{informaticadatacatalogue} as some commercial examples. A more
general approach to data discovery is Aurum~\cite{aurum}, which provides most of
the functionality required to implement the systems above.

\mypar{Data Integration} Relevant work in data integration 
is query reverse engineering~\cite{reversequery1, reversequery2}, as
well as query-by-example interfaces to data integration
such as S4~\cite{s4}. Modern data integration systems such as
Civilizer~\cite{civilizer} and BigGorilla~\cite{biggorilla} assume the existence
and participation of a human expert that needs to build DAGs of integration
operators during the integration activity. \update{R4D10}{Before these
platforms, the}
\emph{Dataspaces}~\cite{dataspaces} \update{}{vision outlined many of the challenges and
opportunities that we still demand of an integration system. We borrowed the
term data mashup from the Yahoo Pipes system}~\cite{yahoopipes}. Related to
creating mashups given many different datasets, some work~\cite{lessismore12}
has studied the diminishing returns of integrating datasets.

\mypar{Data Fusion and Truth Discovery} Data fusion refers to the ability to
combine multiple sources of information to improve the quality of the end
result. In our vision, we consider data fusion operators that permit combining
multiple (possibly diverging) datasets and offer the result to users. This can
be useful, among others, for truth discovery~\cite{truthdiscovery}: the process
of identifying the real value for a specific variable.  The database community
has contributed results to these areas~\cite{srivbdi13, truthfinding1,
truthdiscovery2, truthdiscovery3}. We are building on top of this work to design
fusion operators that can be incorporated into the DMMS architecture.

The DMMS systems we envision aim to incentivize buyers and sellers to solve the
lack of information and incentives problem that keeps data siloed within and
across organizations. At the same time, all the work above is relevant to build
the mashup builder, which is one piece of the larger class of DMMS systems we
envision.

\section{Discussion}
\label{sec:discussion}

In this paper, we presented a vision for data market platforms that focus on the
problems of data sharing, discovery, and integration. These problems are a main
hurdle to organizations' ability to exploit data.

\smallskip

\mypar{Understanding data} While data and artificial intelligence are driving
many changes to our economic, social, political, financial, and legal systems,
we know surprisingly little about their foundations and governing dynamics.
Furthermore, to an extent unseen in previous economic upheavals, the rapid pace
of technological and social innovation is straining the ability of policy and
economic practice to keep up. Moreover, while the recombination and integration
of diverse data creates vast new value, we currently have neither theory for how
data can be combined nor industrial policy for how to protect against the
personal exposures and abuses that grow in proportion. We remain stuck with old
models for understanding these new phenomena and antiquated heuristics for
making decisions in the face of change. The data markets we propose are a
vehicle to initiate the study of theory and systems to address this challenge.

\smallskip

We expect that the insights, algorithms, and systems we will produce will inform
the design of future data market platforms. We expect that the different
systems, simulators, and approaches proposed will pose interesting research
questions for the database community.

\bibliographystyle{abbrv}
\balance
\bibliography{main}

\end{document}